\documentstyle[twocolumn,epsf]{article}
\begin{document}
\twocolumn[%
\vspace{1cm}
\centerline{\Large\bf Paired and Stripe States in the Quantum Hall System} 

\ 

\centerline{Nobuki Maeda}

\ 

\centerline{
Department of Physics, Hokkaido University, 
Sapporo 060-0810, Japan}

\ 

\centerline{
\begin{minipage}{14cm}
{\small We study a paired state at the half-filled Landau level using a 
mean field theory on the von Neumann lattice. 
We obtain a microscopic model which shows a continuous transition 
from the compressible stripe state to the paired state. 
The energy gap in the paired state is calculated numerically 
at the half-filled second Landau level. }
\end{minipage}
}

\ 

\ 

Keywords:  Quantum Hall system, BCS theory, Charge density wave, 
von Neumann lattice

\ 

]

\vspace{3cm}

In recent experiments, highly anisotropic states are observed in the 
half-filled third and higher Landau levels\cite{stripe}. 
In the half-filled second Landau level the fractional quantum Hall 
effect (FQHE)\cite{d} and transition to the highly anisotropic state are 
observed\cite{o}. 
Theoretically the anisotropic states are regarded as stripe states\cite{m} 
or unidirectional charge density wave states which are plausible in the 
Hartree-Fock approximation and numerical calculation of 
small systems\cite{k,n}. 
On the other hand, the FQHE state at the half-filled second Landau level 
is regarded as a paired state like a superconducting state\cite{e,f,g,gg}. 
However the microscopic mechanism of the pairing is not understood yet. 
How is the paired state formed by the Coulomb interaction effect? 
Naively it seems that there is no possibility of formation of the 
paired state by the repulsive force in the mean field theory. 

In the present work, we analyze the gap equation for the Coulomb 
interaction which is screened in the Landau level space 
and find a possibility of transition from the stripe state to a 
paired state by varying the screening length. 
We use a basis on the von Neumann lattice which is useful for the 
field theoretical study of the quantum Hall system\cite{i,j}. 
It was shown that the von Neumann lattice formalism is very 
useful tool for the anisotropic compressible state at the higher 
Landau levels\cite{k,i}. 
The effective potential on the von Neumann lattice becomes 
attractive for a small screening length which is on the order 
of the magnetic length. 
We present a microscopic model which shows a continuous transition 
from the compressible stripe state to the paired state. 

Let us consider two-dimensional electron systems in the presence of a 
perpendicular uniform magnetic field $B$. 
The free particle energy is quenched to the Landau level energy 
$E_l=\hbar\omega_c (l+1/2)$, $l=0,1,2\dots$, where $\omega_c=eB/m$. 
We suppose that the electrons are spin-polarized and ignore 
the spin degree of freedom. 
For simplicity, we set $\hbar=c=1$. 
The total Hamiltonian of the present system is 
\begin{eqnarray}
H&=&\int d^2r\psi^\dagger({\bf r}){(-i\nabla+eA)^2\over2m}\psi({\bf r})\\
&&+{1\over2}\int d^2rd^2r':\delta\rho({\bf r})V({\bf r}-{\bf r}')
\delta\rho({\bf r}'):,
\nonumber
\end{eqnarray}
where colons mean the normal ordering, $\nabla\times A=B$, 
$\delta\rho({\bf r})=\psi^\dagger({\bf r})\psi({\bf r})-\rho_0$, 
$V({\bf r})=q^2/r$, and $\rho_0$ is the uniform background density. 

We project the system to the $l$ th Landau level space. 
The electronic field $\psi$ is expanded by the Wannier basis 
$w_{l,X}({\bf r})$ on the von Neumann lattice as
\begin{equation}
\psi({\bf r})=\sum_{l,X}b_l({\bf X})w_{l,X}({\bf r}).
\end{equation}
The lattice spacing is given by $a=\sqrt{2\pi\hbar/eB}$. 
We set $a=1$ for simplicity. 
${\bf X}$ is defined on the lattice sites. 
$b_l$ is the anti-commuting annihilation operator. 
The system is translationally invariant in the lattice space and 
in the momentum space\cite{i}. 
The translational symmetry in the momentum space is called the K-invariance. 

We apply the mean field approximation to the projected Hamiltonian. 
Let us consider the following mean fields, 
which are translationally invariant on the von Neumann lattice, 
\begin{eqnarray}
U_l({\bf X}-{\bf X}')&=&\langle b_l^\dagger({\bf X}')b_l({\bf X})\rangle,
\nonumber\\
U_l^{(+)}({\bf X}-{\bf X}')&=&\langle b_l^\dagger({\bf X}')b_l^\dagger
({\bf X})\rangle,\\
U_l^{(-)}({\bf X}-{\bf X}')&=&\langle b_l({\bf X}')b_l({\bf X})\rangle.
\nonumber
\end{eqnarray}
These mean fields break the K-invariance and U(1) symmetry. 
Using these mean fields, we obtain a mean field Hamiltonian 
\begin{eqnarray}
H^{(l)}_{\rm mean}&=&\sum_{XX'}[\varepsilon_{l,{\bf X}-{\bf X}'}
b_l^\dagger({\bf X})b_l({\bf X}')
\nonumber\\&&+
{1\over 2}\Delta^*_{l,{\bf X}'-{\bf X}}b({\bf X})b({\bf X}')\\&&+
{1\over 2}
\Delta_{l,{\bf X}'-{\bf X}}b^\dagger({\bf X}')b^\dagger({\bf X})].
\nonumber
\end{eqnarray}
This Hamiltonian is diagonalized by the Bogoliubov transformation in the 
momentum space. 
The gap potential $\Delta_l(p)$ has four zeros at ${\bf p}=(0,0)$, 
$(0,\pi)$, $(\pi,0)$, and $(\pi,\pi)$. 
We assume $\varepsilon_l(-p)=\varepsilon_l(p)$. 
$U_l$ and $U_l^{(-)}$ are calculated by using $H^{(l)}_{\rm mean}$. 
The gap equation is given by 
\begin{equation}
\Delta_l(p)=-\int{d^2p_1\over(2\pi)^2}{\Delta_l(p_1)\over2
E_l(p_1)}{\tilde v}_l(p_1-p)e^{i\int_p^{p_1}
2\alpha(k)dk},
\label{self}
\end{equation}
where $\mu$ is the chemical potential and $E_l(p)$ is the 
spectrum of the quasiparticle which is defined by 
$E_l(p)=\sqrt{\xi_l(p)^2+\vert\Delta_l(p)\vert^2}$, 
$\xi_l(p)=\varepsilon_l(p)-\mu$. 
Note that the gauge field appears in the gap equation (\ref{self}). 
The gauge field represents two unit flux on the Brillouin zone. 
The filling factor of the $l$ th Landau level is denoted by $\nu_l$ and 
total filling factor is given by $\nu=l+\nu_l$. 

Next, we analyze gap equation (\ref{self}). 
It is shown that the screening effect plays important roles 
for the pairing mechanism. 
For simplicity, we omit the Landau level index $l$ and use $q^2/a$ as 
the unit of energy in the following. 

Eq.~(\ref{self}) is rewritten as 
\begin{equation}
\Delta(p)=-\int{d^2k\over(2\pi)^2}{\tilde v}(k) e^{ik\cdot D}{\Delta(p)
\over 2E(p)},
\end{equation}
where ${\bf D}=(-i{\partial\over\partial p_x}+2\alpha_x,
-i{\partial\over\partial p_y}+2\alpha_y)$. 
It is convenient to introduce eigenfunctions of the operator ${\bf D}^2$, 
that is, ${\bf D}^2 \psi_n(p)=e_n \psi_n(p)$ with $e_n=(2 n+1)/\pi$, 
$n=0,1,2,\dots$. 
The index $n$ labels the $n$ th Landau level in the momentum space. 
The eigenfunctions are doubly degenerate and are given by
\begin{eqnarray}
\psi^{(2)}_n(p)
&=&{2\pi\over\sqrt{n!}}({\pi\over2})^{n/2}(D_x-iD_y)^n
\psi^{(2)}_0(p),\\
\psi^{(3)}_n(p)
&=&{2\pi\over\sqrt{n!}}({\pi\over2})^{n/2}(D_x-iD_y)^n \psi^{(3)}_0(p),
\end{eqnarray}
where $N_n=1/\sqrt{2^{n-1}n!}$, and 
$\psi^{(2)}_0$, $\psi^{(3)}_0$ are written by theta functions as 
\begin{eqnarray}
\psi^{(2)}_0(p)&=&N_0 e^{-{p_y^2\over2\pi}}
\vartheta_2({p_x+ip_y\over\pi}\vert 2i),
\label{ptwo}\\
\psi^{(3)}_0(p)&=&N_0 e^{-{p_y^2\over2\pi}}
\vartheta_3({p_x+ip_y\over\pi}\vert 2i).
\label{pthree}
\end{eqnarray}
These eigenfunctions have the parity symmetry, $\psi_n(-p)=(-)^n\psi_n(p)$. 
Therefore the gap potential can be expanded by $\psi_{2 n+1}(p)$. 
We can expand the gap potential as
\begin{eqnarray}
\Delta(p)&=&\sum_{n\geq1,i=2,3}c_n^{(i)}\psi_{2 n-1}^{(i)}(p),
\label{exde}\\
{\Delta(p)\over E(p)}&=&\sum_{n\geq1,i=2,3}d_n^{(i)}\psi_{2 n-1}^{(i)}(p).
\end{eqnarray}
Then Eq.~(\ref{self}) becomes $c_n^{(i)}=-{F_{2 n+1}\over2}d_n^{(i)}$ 
where $F_n$'s are given by the same form as the Haldane's 
pseudo-potentials for the $l$-th Landau level. 

The Coulomb potential is screened by the polarization 
$\Pi(p)$ due to the Fermion loop diagrams. 
We approximate the screened potential as
\begin{equation}
{\tilde v}(p,m_{\rm TF})=1/({\tilde v}(p)^{-1}+m_{\rm TF}),
\end{equation}
where $m_{\rm TF}=-\Pi(0)$ is the Thomas-Fermi mass. 
Calculating the one-loop diagram, $m_{\rm TF}$ is given by 
\begin{equation}
m_{\rm TF}=\int_{\bf BZ}{d^2p\over(2\pi)^2}{\vert\Delta(p)\vert^2
\over 2E(p)^3}
\label{tf}
\end{equation}
We use the screened potential ${\tilde v}(p,m_{\rm TF})$ in the 
gap equation. 
We also use the screened potential to calculate $\varepsilon_l$ 
in the Hartree-Fock approximation. 

The $m_{\rm TF}$ dependences of pseudopotentials $F_n$ are plotted 
in Fig.~(1) for $l=1$. 
As seen in the figure, potentials change their signs at 
$m_{\rm TF}\approx 1.0$. 
Thus, in the Hartree-Fock-Bogoliubov approximation, 
the stripe phase makes transition to the pairing phase 
for $m_{\rm TF}>1.0$. 

\begin{figure}
\centerline{
\epsfysize=1.8in\epsffile{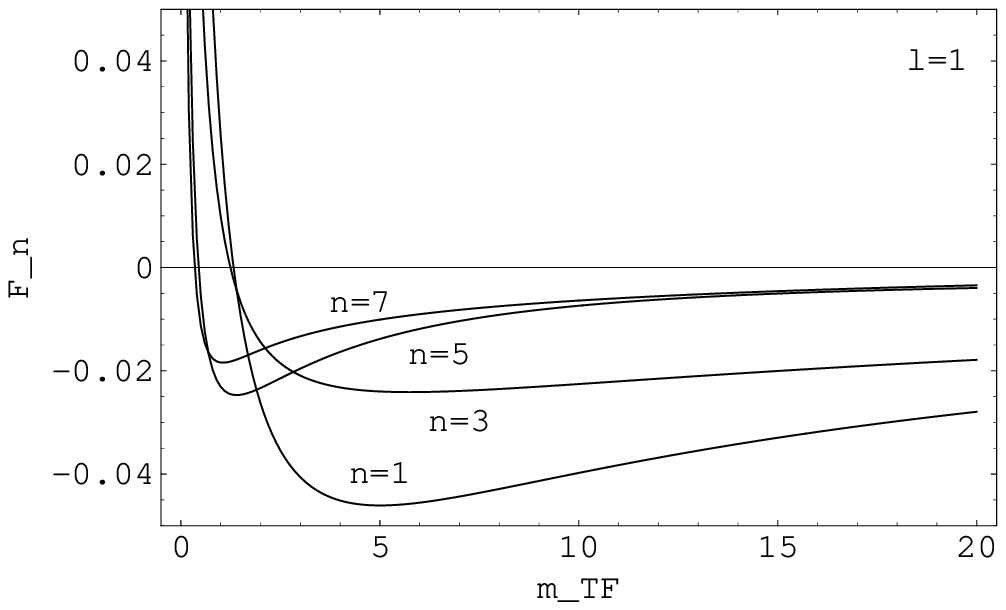}}
Fig~1. The $m_{\rm TF}$ dependence of the pseudo-potentials for 
$l=1$. The unit of the energy is $q^2/a$. 
\end{figure}

We present an effective Hamiltonian which 
has a striped state as the normal state and paired state as the 
U(1) symmetry breaking state. 
We focus our argument on the half-filled second Landau level 
space, that is $l=1$, $\nu_1=1/2$. 

We truncate the expansions of Eq.~(\ref{exde}) and 
calculate the self-consistent solution by iteration of numerical 
calculations until we obtain convergence. 
Considering only the lowest and next relevant terms, the effective 
Hamiltonian for the quasiparticle in the striped and paired state 
is given by
\begin{eqnarray}
H_{\rm eff}&=&\int_{\rm BZ}{d^2p\over(2\pi)^2}[
\varepsilon_{\rm eff}(p)
a^\dagger(p) a(p)\nonumber\\
&&+{1\over2}\Delta_{\rm eff}(p) 
a^\dagger(-p) a^\dagger(p)
\nonumber\\
&&+{1\over2}\Delta^*_{\rm eff}(p) a(p)a(-p)],
\end{eqnarray}
where the effective hopping potential and effective gap potential 
are given by 
\begin{eqnarray}
\varepsilon_{\rm eff}(p)&=&-t_{\rm eff}\cos p_y-t_{(0,3)}\cos 3 p_y,
\label{epe}\\
\Delta_{\rm eff}(p)&=&c_1^{(3)}\psi_1^{(3)}(p)+c_3^{(3)}\psi_3^{(3)}(p). 
\label{dele}
\end{eqnarray}
$a(p)$ is the anti-commuting annihilation operator in the momentum space. 
The hopping parameters $t_{(0,3)}$ and gap potential $\Delta_{\rm eff}(p)$ 
depend on $t_{\rm eff}$. 
The magnitude of $t_{\rm eff}$ corresponds to the strength of the 
stripe order. 

Chemical potential $\mu$ is determined so that the filling factor 
$\nu_1$ is equal to $1/2$. 
Note that $\mu$ includes the on-site term in $H_{\rm mean}$. 
We find that $\mu$ is negative small number on the order of $10^{-4}$ 
at most. 
The maximum value of the energy gap is $0.027q^2/a=0.01q^2/l_B 
(l_B=\sqrt{\hbar/eB})$ at $t_{\rm eff}=0.03$. 
This value is the same order as Morf's one\cite{e}. 
The excitation energy $E(p)$ becomes small around $p_y=\pm\pi/2$, 
which is the Fermi surface of the striped state. 
The transition to the stripe phase is continuous and very smooth. 
Near the transition point, the behavior of energy gap is approximated by  
$t_{\rm eff} e^{-2\pi t_{\rm eff}/\vert F_1\vert}$, whose non-perturbative 
dependence on the coupling is well-known in the BCS theory. 
At $m_{\rm TF}= m_c\approx 1.4$, $F_1$ behaves as $\alpha (m_c-m_{\rm TF})$ 
and the energy gap approaches to zero at $t_{\rm eff}=t_c\approx0.2$. 
The energy gap is extremely small at $0.1<t_{\rm eff}<t_c$. 
At $t_{\rm eff}>t_c$, the gap potential vanishes and the compressible 
striped state is realized. 
This state leads to the anisotropy of the magnetoresistance
\cite{k}. 
Inspecting the energy spectrum of quasiparticle in the pairing phase, 
we find a crossover phenomenon at $t_{\rm eff}\approx0.01$, 
that is, the minimum excitation energy $\min(E(p))$ is placed around 
$p_y=\pm\pi/2$ at $0.01<t_{\rm eff}<t_c$, 
whereas at $0<t_{\rm eff}<0.01$, placed around $p_y=0$ and 
$\pi$. 
We call the latter case the gap-dominant pairing phase. 
In this phase, the low-energy excitation occurs around zeros of the 
gap potential and the the energy gap is given by $2\xi(0)$. 
The spectrum is close to the flatband and 
the stripe order is weakened. 
The $t_{\rm eff}$ dependence of the energy gap is summarized as
\begin{equation}
\Delta E\propto\left\{
\begin{array}{ll}
t_{\rm eff}&{\rm for}\  0<t_{\rm eff}<0.01,\\
t_{\rm eff} e^{-{2\pi t_{\rm eff}/\vert F_1\vert}}
&{\rm for}\  0.1<t_{\rm eff}<t_c,\\
0& {\rm for}\  t_c<t_{\rm eff}.
\end{array}
\right.
\end{equation}

In conclusion, we obtain a microscopic model which shows a continuous 
transition from stripe to paired state. 
The gap potential has the p-wave-like pairing as seen in Eq.~(11). 
Details of the calculation and numerical results are given 
in Ref.~\cite{p}. 
More experimental and theoretical studies are necessary for deciding the 
symmetry of the gap potential and understanding the underlying physics 
at the half-filled second Landau level. 

This work was partially supported by the special Grant-in-Aid for 
Promotion of Education and Science in Hokkaido University provided by the 
Ministry of Education, Science, Sport, and Culture, the Grant-in-Aid for 
Scientific Research on Priority area (Physics of CP violation) 
(Grant No. 12014201), and the Grant-in aid for International Science 
Research (Joint Research 10044043) from the Ministry of Education, 
Science, Sports, and Culture, Japan.

\end{document}